# Theory of coherence in Bose-Einstein condensation phenomena in a microwave driven interacting magnon gas

Sergio M. Rezende

*Departamento de Física, Universidade Federal de Pernambuco, Recife, PE 50670-901, Brazil*

Strong experimental evidences of the formation of quasi-equilibrium Bose-Einstein condensation (BEC) of magnons at room temperature in a film of yttrium iron garnet (YIG) excited by microwave radiation have been recently reported. Here we present a theory for the magnon gas driven by a microwave field far out of equilibrium showing that the nonlinear magnetic interactions create cooperative mechanisms for the onset of a phase transition leading to the spontaneous generation of quantum coherence and magnetic dynamic order in a macroscopic scale. The theory provides rigorous support for the formation of a BEC of magnons in a YIG film magnetized in the plane. We show that the system develops coherence only when the microwave driving power exceeds a threshold value and that the theoretical result for the intensity of the Brillouin light scattering from the BEC as a function of power agrees with the experimental data. The theory also explains quantitatively experimental measurements of microwave emission from the uniform mode generated by the confluence of BEC magnon pairs in a YIG film when the driving power exceeds a critical value.

## I. Introduction

In a recent series of papers Demokritov and co-workers have reported remarkable experimental evidence of the formation of Bose-Einstein condensation (BEC) and related phenomena in a magnon gas driven by microwave radiation [1-6]. Bose-Einstein condensation, a phenomenon that occurs when a macroscopic number of bosons occupies the lowest available quantum energy level [7], has only been unequivocally observed in a few physical systems, such as superfluids [7], excitons and biexcitons in semiconductors [8,9], atomic gases [10] and certain classes of quantum magnets [11]. BEC phenomena usually takes place by cooling the system to very low temperatures. The room temperature experiments reported in [1-6] have ingeniously materialized earlier proposals for producing Bose-Einstein condensation of magnons [12,13] and demonstrated powerful techniques for observing its unique properties.

The experiments were done at room temperature in epitaxial crystalline films of yttrium-iron garnet (YIG) magnetized by an applied in-plane field. In these films the combined effects of the exchange and magnetic dipolar interactions among the spins produce a dispersion relation (frequency $\omega_k$ versus wavevector $k$) for magnons propagating at angles with the field smaller than a critical value that has a minimum $\omega_{k_0}$ at $k_0 \sim 10^5$ cm$^{-1}$. In bulk samples the dispersion relation has the usual parabolic shape with a minimum at $k = 0$, where the density of states vanishes. In films the energy minimum away from the Brillouin zone center produces a peak in the density of states at $\omega_{k_0}$, providing an important condition for the formation of the condensate.

The experiments reported in [1-6] employ a microwave magnetic field with pumping frequency $f_p$ = 8.1 GHz applied parallel to the static field in the so-called parallel pumping process [14,15] to drive magnons in YIG films magnetized in the plane. In some of the latest experiments reported [4,5], short microwave pulses (30 ns) are used to create a hot magnon gas, allowing its evolution to be observed with time resolved Brillouin light scattering (BLS). Several important features are observed with increasing microwave power. Initially, when the power exceeds a first threshold value, there is a large increase in the population of the parametric magnons with frequency in a narrow range around $f_p/2$ = 4.05 GHz. Then the energy of these primary magnons redistributes in about 50 ns through modes with lower frequencies down to the minimum frequency $f_{min} = \omega_{k0}/2\pi$ = 2.9 GHz (for $H$ = 1.0 kOe) as a result of magnon interactions that conserve the number of magnons. This produces a hot magnon gas that remains decoupled from the lattice for over 200 ns due to the long spin-lattice relaxation time. The BLS spectrum in this time span reflects the shape of the magnon density of states weighted by the appropriate thermal distribution exhibiting a peak at the frequency $f_{min}$. Thereafter this peak decays exponentially in time in the range of several hundred ns due to the thermalization with the crystal lattice. However if the microwave power exceeds a second threshold value, much larger than the one for parallel pumping, two striking features are observed, namely, the decay rate of the BLS peak at $f_{min}$ doubles in value while its intensity increases by two orders of magnitude. The behavior of the BLS peak was attributed to a change in the magnon state from incoherent to coherent, indicating the formation of a room-temperature BEC of magnons [4,5].

Coherence of photon fields has a formal quantum treatment developed by Roy J. Glauber over four decades ago [16]. Coherent magnon states, introduced in analogy with the photon states also have a formal quantum treatment [17,18]. In this paper we show that an interacting magnon system in a YIG film driven by microwave radiation develops a spontaneous coherent state with properties that explain the main features of the experimental observations. Since the coherent state corresponds to a quantum macroscopic wavefunction, the theory provides rigorous support for the existence of Bose-Einstein condensation of magnons in the experiments of [1-6]. Note that recently it has been argued [19] that the intermagnon interactions in a YIG film magnetized in the plane prevents the conditions for stabilization of the BEC. Contrary to the conclusions of [19], we show that the magnon-magnon interactions play an essential role in the formation of the BEC at room-temperature in a YIG film driven by microwave radiation as in the experiments of Demokritov and co-workers [1-6].

In another recent paper of the same group, Dzyapko *et al.* [6] show that if the applied in-plane static field has a value such that the frequency of the $k \approx 0$ magnon is $\omega_0 = 2\omega_{k0}$, a microwave radiation signal is generated by $k \approx 0$ magnons created by pairs of BEC magnons $k_0, -k_0$ through a three-magnon confluent process. The $k \approx 0$ value is necessary for emission because the wavenumber of electromagnetic radiation with frequency 1.5 GHz, as in the experiments [6], is $k = 2\pi/f \approx 0.3$ cm$^{-1}$. In an earlier paper [20] we have shown that the $k \approx 0$ magnons created by the BEC are coherent magnons states, corresponding to a nearly uniform magnetization precessing with frequency $\omega_0$ and generating a microwave signal. The microwave emission from the collective action of the spins is identified with superradiance. Here we present other aspects of the theoretical model for this phenomenon and show that the predicted radiated signal power agrees with the experimental data [6].

The paper is organized as follows. In Sec. II we discuss the nature of the spin-wave modes in thin films based on the results of earlier work by several authors, in order to establish the background for the remainder of the paper. Sec. III is devoted to a review of the properties of coherent magnons states. In Sec. IV we discuss the excitation of spin waves in films by the parallel pumping technique. Sec. V is devoted to the proposed cooperative mechanism for the formation of the BEC of magnons. In Sec. VI we show that the states resulting from the cooperative action have quantum coherence. In Sec. VII we show that the results of the model agree with experimental data for the intensity of the Brillouin light scattering from the BEC and for the microwave emission from the uniform mode resulting from the coalescence of a pair of BEC magnons. Sec. VIII summarizes the main results.

## II. Spin-wave modes in thin films

Since the pioneering work of Damon and Eshbach [21] the theory of spin waves in ferromagnetic films has been studied and reviewed by many authors [22-31]. The theory of Damon and Eshbach (DE) was developed for waves with very small wavenumbers *k* that have energies with negligible contribution from the exchange interaction between the spins. They used a semi-classical approach for the equation of motion for the magnetization in which the magnetic dipolar field plays a dominant role. This field was obtained with the so-called magnetostatic approximation valid for wavenumbers much larger than the values for the electromagnetic field ($k \sim 1$ cm$^{-1}$) so DE coined the term magnetostatic waves to the resulting wave solutions. Later several authors included the exchange interaction in the equations of motion and in the boundary conditions, used various approaches and approximations to find the normal modes and introduced other names to the waves, such as dipole-exchange waves. Actually they are all simply spin waves, pictured classically by the view of the spins precessing about the equilibrium direction with a phase that varies along the direction of propagation. Various results have been successfully applied to explain experimental observations in thin slabs or films of YIG and other low loss ferrite materials as well as in ultrathin films of ferromagnetic metals. In this section we present the background information on the normal spin-wave modes in a thin ferromagnetic film necessary for the discussion of the theory of the interacting spin waves. Initially we employ the DE theory extended to include exchange in order to obtain exact dispersion relations for waves in films corresponding to the nearly uniform transverse mode. Then we develop a quantum model based on the second quantization of the spin excitations involving magnon creation and annihilation operators which is the most convenient approach to treat interactions.

Consider an unbounded flat ferromagnetic film with thickness *d* magnetized in the plane by a static magnetic field *H*. We use a coordinate Cartesian system with the *x* and *z* coordinates in the plane of the film, $\hat{z}$ along the field direction. Anisotropy is neglected since it is very small in YIG so that the magnetization $\vec{M}$ in equilibrium lies along $\hat{z}$ and one can write $\vec{M} = \hat{z} M_z + \hat{x} m_x + \hat{y} m_y$. The DE approach consists of solving the Landau-Lifshitz equations of motion for the small-signal time-varying components of the magnetization $m_x$ and $m_y$ under the action of the magnetic dipolar field they create added to the static field *H* [21]. Furthermore it is assumed that $m_x$ and $m_y$ are described by waves with frequency $\omega_k$ and wavevector $\vec{k}$ propagating in the film *x-z* plane and a standing wave pattern in the perpendicular direction. The corresponding dipolar field $\vec{h}_d$ can be obtained from Maxwell's equations in the magnetostatic approximation $\nabla \times \vec{h}_d = 0$, which allows expressing the field in terms of a magnetic potential $\psi$ as $\vec{h}_d = -\nabla \psi$. The equation for $\psi$ follows from $\nabla \cdot (\vec{h}_d + 4\pi \vec{m}) = 0$ and its solutions are subjected to the electromagnetic boundary conditions involving the internal and external fields on the two surfaces of the film. One then obtains a transcendental equation relating the frequency $\omega_k$ with the wavevector components [21,27-29],

$$2(1+\kappa)(-\delta)^{1/2} \cot(k_y d) + \delta(1+\kappa)^2 - \nu^2 \sin^2\theta_k + 1 = 0 \qquad (1)$$

where $\theta_k$ is the angle between the wavevector $\vec{k}$ in the plane and the *z*-direction, $k_y$ is the wavenumber characterizing the mode pattern in the direction normal to the film and the other parameters are related to the frequency by

$$\kappa = \frac{\omega_H \omega_M}{\omega_H^2 - \omega_k^2} \quad , \quad \nu = \frac{\omega_k \omega_M}{\omega_H^2 - \omega_k^2} \qquad \text{and} \qquad (2)$$

$$\delta = \frac{1+\kappa \sin^2\theta_k}{1+\kappa} \quad , \qquad (3)$$

where $\omega_H = \gamma H$, $\omega_M = \gamma 4\pi M$ and $\gamma = g\mu_B/\hbar$ is the gyromagnetic ratio (2.8 GHz/kOe for YIG). Note that the components of the wavevector $\vec{k}$ in the plane enter in (1)-(3) through $k_x = k \sin\theta_k$ and $k_z = k \cos\theta_k$.

From the equation for the potential in the film one can see [21,29] that the transverse wavenumber $k_y$ is related to the wavenumber $k$ in the plane by

$$k_y = (-\delta)^{1/2} k \ . \tag{4}$$

It follows that for each pair of values of $k_x$, $k_z$, or equivalently $k$, $\theta_k$, Equation (1) has several solutions for the frequency $\omega_k$, each corresponding to a different tranverse mode pattern characterized by a discrete $k_y$. From (2)-(4) it is clear that $k_y$ can be real or imaginary, depending on the range of frequency. Real values of $k_y$ correspond to the so-called volume magnetostatic modes, for which the magnetization components have a dependence on the transverse coordinate $y$ of the type cos $k_y y$, sin $k_y y$. Imaginary values of $k_y$ correspond to the surface modes, which have an exponential dependence on $y$ decaying away from one of the film surfaces. The surface modes have a unique property of being non-reciprocal, in the sense that the wave associated with one surface propagates only in one direction but not in the opposite [21,22,27-29]. From (2)-(4) it can be shown [27-29] that for each frequency there is a critical angle of propagation $\theta_{kc}$ above which $\delta$ becomes positive so that $k_y$ is imaginary and the mode is a surface wave,

$$\sin \theta_{kc} = (\frac{\omega_k^2 - \omega_H^2}{\omega_H - \omega})^{1/2} \ . \tag{5}$$

For typical numbers appropriate to the experiments of [1-6] with YIG films, $H = 1.0$ kOe, $4\pi M = 1.76$ kG, $f = \omega_k / 2\pi = 4.0$ GHz, the critical angle is $\theta_{kc} = 50.26°$. For the specific case of the surface wave with $\theta_k = 90°$, Equation (1) has a simple solution with an explicit dependence of the frequency on the wavevector given by [29],

$$\omega_k^2 = \omega_H^2 + \omega_H \omega_M + \frac{1}{4} \omega_M^2 (1 - e^{-2kd}) \ . \tag{6}$$

The introduction of the exchange interaction complicates considerably the problem of finding the spin-wave normal modes in films. First of all one can see that in films with thickness on the order of 1 μm or less, the exchange introduces a sizeable separation in the frequencies of the volume modes with different transverse patterns because $k_y \sim n_y \pi / d$ and the exchange energy varies with the square of $k_y$. The exact solution of the wave equations must involve the matching of mixed electromagnetic and exchange boundary conditions [24-26]. A nearly exact expression for the frequency of the lowest lying exchange branch can be obtained by simply introducing the exchange interaction as an effective field in Equations (1)-(4) which is added to the applied field, so that the parameter $\omega_H$ becomes,

$$\omega_H = \gamma (H + D k^2) \ , \tag{7}$$

where $D = 2 J S a^2 / g \mu_B$ is the exchange stiffness, $J$ being the nearest neighbor exchange constant and $a$ the lattice parameter of the film. The dispersion relations obtained by solving numerically Equations (1)-(4), with $\omega_H$ as in (7), are shown by the solid lines in Figure 1 for several angles $\theta_k$ in two YIG films with thickness $d = 0.1$ μm and 5 μm, using $H = 1.0$ kOe, $4\pi M = 1.76$ kG, and $D = 2 \times 10^{-9}$ Oe.cm$^2$. The main feature of the dispersion curves is that for propagation angles below certain values the frequency exhibits a minimum at a $k$ value that depends on the thickness. This is a consequence of the fact that the frequency initially decreases with increasing $k$ due to the role of the dipolar energy but then at larger values of $k$ it changes slope due to the effect of exchange as in (7).

In the quantum approach which will be used to treat interactions we use a Hamiltonian in the form,

$$H = H_0 + H_{int} + H'(t) \ , \tag{8}$$

where $H_0$ is the unperturbed Hamiltonian that describes free magnons, $H_{int}$ represents the nonlinear magnetic interactions and $H'(t)$ represents the external microwave driving. The magnetic Hamiltonian can be written as $H = H_Z + H_{exc} + H_{dip}$, representing respectively, the Zeeman, exchange, and dipolar contributions. We treat the quantized excitations of the magnetic system with the approach of Holstein-Primakoff [32-35], which consists of three transformations that allow the spin operators to be expressed in terms of boson operators that create or destroy magnons. In the first transformation the components of the local spin operator are related to the creation and annihilation operators of spin deviation at site $j$, denoted respectively by $a_j^+$ and $a_j$, which satisfy the boson commutation rules $[a_i, a_j^+] = \delta_{ij}$ and $[a_i, a_j] = 0$. Using a

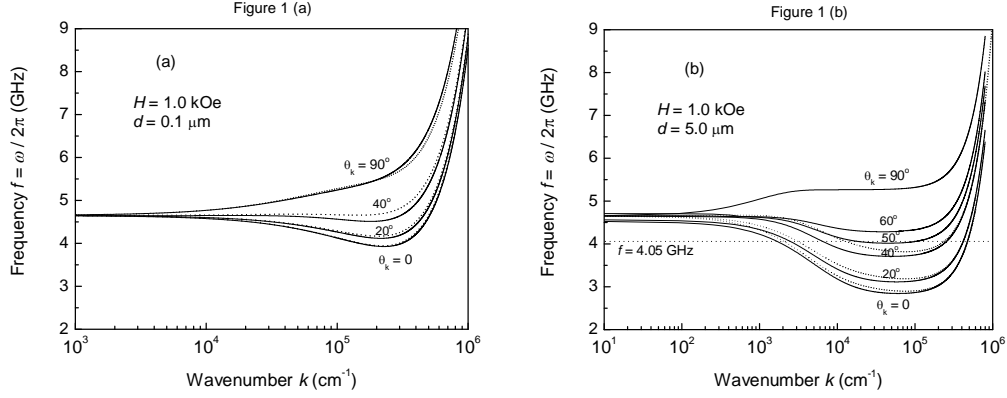

Figure 1: Dispersion relations for spin waves propagating at various angles with the in-plane applied field $H = 1.0$ kOe in a YIG film with thickness (a) 0.1 μm and (b) 5 μm. The curves with full lines represent the calculation with the DE theory including exchange, Equation (1) while the dotted lines represent the calculation with the approximate theory Equation (26) for $\theta_k = 0$, 20° and 40°.

coordinate system with $\hat{z}$ along the equilibrium direction of the spins, defining $S_j^+ = S_j^x + iS_j^y$ and $S_j^- = S_j^x - iS_j^y$, where the factor $i$ is the imaginary unit, not to be confused with the subscript denoting lattice site $i$, it can be shown that the relations that satisfy the commutation rules for the spin components and the boson operators are [32-34]

$$S_j^+ = (2S)^{1/2} (1 - \frac{a_j^+ a_j}{2S})^{1/2} a_j, \tag{9a}$$

$$S_j^- = (2S)^{1/2} a_j^+ (1 - \frac{a_j^+ a_j}{2S})^{1/2}, \tag{9b}$$

$$S_j^z = S - a_j^+ a_j, \tag{9c}$$

where $S$ is the spin and $n_j = a_j^+ a_j$ is the operator for the number of spin deviations at site $j$. One of the main advantages of this approach is that the nonlinear interactions are treated analytically by expanding the square root in (9a) and (9b) in Taylor series. We use only the first two terms of the expansion, so that

$$S_j^+ \cong (2S)^{1/2} (a_j - a_j^+ a_j a_j / 4S) \tag{10a}$$

and
$$S_j^- \cong (2S)^{1/2} (a_j^+ - a_j^+ a_j^+ a_j / 4S). \tag{10b}$$

In order to find the normal modes of the system we use the linear approximation, whereby only the first terms in (9c) and (10) are kept, i.e., $S_j^+ \cong (2S)^{1/2} a_j$, $S_j^- \cong (2S)^{1/2} a_j^+$, and $S_j^z \cong S$. With these transformations one can express the magnetic Hamiltonian in a quadratic form containing only lattice sums of products of two boson operators. The second step is to introduce a transformation from the localized field operators to collective boson operators $a_k^+$ and $a_k$ using the Fourier transform

$$a_j = \frac{1}{N^{1/2}} \sum_k e^{i \vec{k} \cdot \vec{r}} a_k \tag{11}$$

where $N$ is the number of spins in the system. The condition that the new collective operators satisfy the boson commutation rules $[a_k, a_{k'}^+] = \delta_{k,k'}$ and $[a_k, a_{k'}] = 0$, requires that the transformation coefficients satisfy the usual orthonormality relations. The contributions from the Zeeman and exchange energies to the Hamiltonian $H_0$ with quadratic form in boson operators can be shown to be [32-35]

$$H_z + H_{exc} = \hbar \sum_k \gamma (H + Dk^2) a_k^+ a_k. \tag{12}$$

The contribution of the dipolar energy to the Hamiltonian can be obtained with approximations valid for the nearly uniform transverse mode, which corresponds to the lowest lying exchange branch with $k_y \sim 0$. Following [30-31] we neglect the variation of the magnetization on the transverse coordinate and work with the averages over $y$,

$$m_{x,y}(x,z;t) = m_{x,y}(\vec{r};t) = \int_{-d/2}^{d/2} \frac{1}{d} m_{x,y}(x,z;t)\, dy \quad . \tag{13}$$

The magnetic potential $\psi$ created by the spatial variation of the small-signal transverse components of the magnetization is written in the form [30,31],

$$\psi(x,y,z) = \frac{1}{V^{1/2}} \sum_k \psi_k(y) e^{i\vec{k}\cdot\vec{r}} \tag{14}$$

where $V$ is the volume of the film and $\vec{k}$ and $\vec{r}$ denote the wavevector and the position vector in the plane. The Fourier transform of the potential $\psi_k(y)$ can be obtained from the solution of $\nabla^2 \psi = 4\pi \nabla \cdot \vec{m}$ derived from Maxwell's equations subject to the electromagnetic boundary conditions at $y = \pm d/2$ [31],

$$\psi_k(y) = 4\pi i [e^{-kd/2} \cosh(ky) - 1] \frac{k_x}{k^2} m_x(k) + 4\pi e^{-kd/2} \sinh(ky) \frac{1}{k} m_y(k) \tag{15}$$

where the Fourier components of the magnetization appearing in (15) can be expressed in terms of the collective boson operators using the relation $m_{x,y} = g\mu_B (N/V) S_{x,y}$ in (10) and (11),

$$m_x(k) = \hbar\gamma \left(\frac{NS}{2V}\right)^{1/2} (a_k + a_{-k}^+) \quad , \tag{16a}$$

$$m_y(k) = -i\hbar\gamma \left(\frac{NS}{2V}\right)^{1/2} (a_k - a_{-k}^+) \quad . \tag{16b}$$

The small-signal transverse components of the dipolar field can be obtained from the magnetic potential with $\vec{h}_d = -\nabla \psi$ so that the contribution of the dipolar energy to the magnetic Hamiltonian can be calculated with

$$H_{dip} = -\frac{1}{2} \int dx\, dy\, dz\, (m_x h_x^{dip} + m_y h_y^{dip}) \quad . \tag{17}$$

The integration in (17) can be performed without difficulty by expressing the magnetization and the dipolar field in terms of their Fourier transforms and using the orthonormality relations. One can show that,

$$H_{dip} = \hbar\gamma 2\pi M \sum_k [(1-F_k)\sin^2\theta_k + F_k] a_k^+ a_k + \frac{1}{2}\{[(1-F_k)\sin^2\theta_k - F_k] a_k a_{-k} + h.c.\} \quad . \tag{18}$$

With (12) and (18) one can write the total Hamiltonian for the free magnon system as,

$$H_0 = \hbar \sum_k A_k a_k^+ a_k + \frac{1}{2} B_k a_k a_{-k} + \frac{1}{2} B_k^* a_k^+ a_k^+ \tag{19}$$

where

$$A_k = \gamma[H + Dk^2 + 2\pi M (1-F_k)\sin^2\theta_k + 2\pi M F_k] \quad , \tag{20a}$$

$$B_k = \gamma[2\pi M (1-F_k)\sin^2\theta_k - 2\pi M F_k] \quad , \tag{20b}$$

$$F_k = (1 - e^{-kd})/kd \quad . \tag{20c}$$

In order to diagonalize the quadratic Hamiltonian it is necessary to introduce new collective boson operators $c_k^+$ and $c_k$ satisfying the commutation rules $[c_k, c_{k'}^+] = \delta_{kk'}$ and $[c_k, c_{k'}] = 0$, related to $a_k^+$ and $a_k$ through the Bogoliubov transformation [32-34]

$$a_k = u_k c_k + v_k c_k^+ \quad , \tag{21a}$$

$$a_k^+ = u_k c_k^+ + v_k^* c_k \quad , \tag{21b}$$

where $u_k^2 - v_k^2 = 1$, as appropriate for a unitary transformation. The coefficients of this transformation must be such that the quadratic Hamiltonian acquires the diagonal form

$$H_0 = \hbar \sum_k \omega_k c_k^+ c_k \quad , \tag{22}$$

because this leads to the Heisenberg equation of motion

$$\frac{dc_k}{dt} = \frac{1}{i\hbar}[c_k, H_0] = -i\omega_k c_k \quad . \tag{23}$$

This equation has stationary solutions of the form $e^{i\omega_k t}$ which assures that $c_k$ is the operator for the normal-mode excitations of the magnetic system. Hence $c_k^+$ and $c_k$ are the creation and annihilation operators for magnons. It can be shown [32-34] that the coefficients of the transformations (21) are given

$$u_k = \left(\frac{A_k + \omega_k}{2\omega_k}\right)^{1/2} \quad , \tag{24a}$$

and

$$v_k = \pm (u_k^2 - 1)^{1/2} = \pm \left(\frac{A_k - \omega_k}{2\omega_k}\right)^{1/2} \quad , \tag{24b}$$

where the sign of $v_k$ in (24b) is the opposite one of the parameter $B_k$ and the frequency $\omega_k$ of the eigenmodes is

$$\omega_k = (A_k^2 - |B_k|^2)^{1/2} \quad . \tag{25}$$

Using the expressions for the parameters in (20) we obtain from (25) an explicit equation for the dependence of the spin-wave frequency on the wavevector in the plane,

$$\omega_k^2 = \gamma^2 [H + Dk^2 + 4\pi M (1 - F_k) \sin^2 \theta_k][H + Dk^2 + 4\pi M F_k] \tag{26}$$

This equation is the same as the one obtained for the lowest lying branch of the "dipole-exchange" modes with more rigorous treatment of the exchange interaction [24-26]. It also agrees with the results of [30,31] in the limit $kd \ll 1$. The dispersion curves shown by the dotted lines in Figure 1 are obtained with (26). The agreement with the Damon-Eshabch result extended to include exchange is quite good for any angle $\theta_k$ in the YIG film with $d = 0.1$ μm. In the case of the film with $d = 5$ μm the agreement is good for $\theta_k < 50^\circ$. The results in Figure 1 shows that the second quantization approach just presented describes quite well the nearly uniform transverse spin-wave mode in films.

To conclude this section we express the components of the magnetization vector operators in terms of the magnon creation and annihilation operators using the relations with the spin operators and the transformations (10), (11) and (21),

$$m_x(\vec{r}) = \frac{M}{(2NS)^{1/2}} \sum_k e^{i\vec{k}\cdot\vec{r}} (u_k + v_k)(c_k + c_{-k}^*) \quad , \tag{27a}$$

$$m_y(\vec{r}) = -i \frac{M}{(2NS)^{1/2}} \sum_k e^{i\vec{k}\cdot\vec{r}} (u_k - v_k)(c_k - c_{-k}^*) \quad , \tag{27b}$$

With these equations one can calculate the expectation values of the magnetization components for any spin excitation in films expressed in terms of the magnon states.

**III- Coherent magnon states**

If the nonlinear interactions are neglected, the spin-wave excitations with wavevector $k$ and frequency $\omega_k$ described by magnon creation and annihilation operators $c_k^+$ and $c_k$ form a system of independent harmonic oscillators, governed by the unperturbed Hamiltonian $H_0 = \hbar \sum_k \omega_k c_k^+ c_k$. The eigenstates $|n_k\rangle$ of this Hamiltonian which are also eigenstates of the number operator $n_k = c_k^+ c_k$ can be obtained by applying integral powers of the creation operator to the vacuum,

$$|n_k\rangle = [(c_k^+)^{n_k}/(n_k!)^{1/2}]|0\rangle, \quad (28)$$

where the vacuum state is defined by the condition $c_k|0\rangle = 0$. These stationary states describe systems with a precisely defined number of magnons $n_k$ and uncertain phase. They form a complete orthonormal set which can be used as a basis for the expansion of any state of spin excitation. They are used in nearly all quantum treatments of thermodynamic properties, relaxation mechanisms, and other phenomena involving magnons. However, as can be seen from the expressions in (27), they have zero expectation value for the small-signal transverse magnetization operators $m_x$ and $m_y$ and thus do not have a macroscopic wavefunction. In order to establish a correspondence between classical and quantum spin waves one should use the concept of coherent magnon states [17,18], defined in analogy to the coherent photon states introduced by Glauber [16]. A coherent magnon state is the eigenket of the circularly polarized magnetization operator $m^+ = m_x + im_y$. It can be written as the direct product of single-mode coherent states, defined as the eigenstates of the annihilation operator,

$$c_k|\alpha_k\rangle = \alpha_k|\alpha_k\rangle, \quad (29)$$

where the eigenvalue $\alpha_k$ is a complex number. Although the coherent states are not eigenstates of the unperturbed Hamiltonian and as such do not have a well defined number of magnons, they have nonzero expectation values for the magnetization $m^+$ with a well defined phase. Here we review a few important properties of the coherent states. First we recall that they can be expanded in terms of the eigenstates of the unperturbed Hamiltonian [16-18],

$$|\alpha_k\rangle = e^{-1/2|\alpha_k|^2} \sum_{n_k} (\alpha_k)^{n_k}/(n_k!)^{1/2}|n_k\rangle. \quad (30)$$

The probability of finding $n_k$ magnons in the coherent state $|\alpha_k\rangle$ obtained directly from (30) is given

$$\rho_{coh}(n_k) = |\langle n_k|\alpha_k\rangle|^2 = (|\alpha_k|^{2n_k}/n_k!)e^{-|\alpha_k|^2}. \quad (31)$$

This function is a Poisson distribution [16] that exhibits a peak at the expectation value of the occupation number operator $\langle n_k\rangle = |\alpha_k|^2$ in the coherent state. It can be shown that coherent states are not orthogonal to one another, but they form a complete set, so that they constitute a basis for the expansion of an arbitrary state. The distribution (31) is very different from the one prevailing in systems in thermal equilibrium, which cannot be described by pure quantum states. Instead they are described by a mixture in which one can find any number of magnons $n_k$ with energy $\hbar\omega_k$. The average number of magnons with energy $\hbar\omega_k$ in thermal equilibrium at a temperature $T$ is given by the Bose-Einstein distribution

$$\bar{n}_k = \frac{1}{e^{\hbar\omega_k/k_B T} - 1} \quad (32)$$

where $k_B$ is the Boltzmann constant. The probability of finding $n_k$ magnons with energy $\hbar\omega_k$ in the mixture describing the thermal equilibrium with the average value (32) can be shown to be [16]

$$\rho_{th}(n_k) = \frac{(\bar{n}_k)^{n_k}}{(1+\bar{n}_k)^{n_k+1}}. \quad (33)$$

Note that for large $n_k$ Equation (33) approaches the exponential function $\exp(-\bar{n}_k)$. To stress the difference between the coherent state and the mixture describing the thermal equilibrium we show in Figure 2 the distributions (31) and (33) corresponding to $\langle n_k\rangle = 50$.

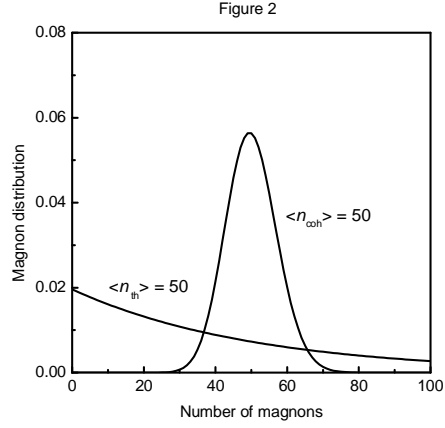

Figure2: Distributions of magnons in a system in thermal equilibrium and in a coherent state with $\langle n_k \rangle = 50$.

Another important property of a coherent state is that it can be generated by the application of a displacement operator to the vacuum [16-18],

$$|\alpha_k\rangle = D(\alpha_k)|0\rangle , \qquad (34a)$$

where

$$D(\alpha_k) = \exp(\alpha_k c_k^+ - \alpha_k^* c_k) . \qquad (34b)$$

In order to study the coherence properties of a magnon system, it is convenient to use the density matrix operator $\rho$ and its representation as a statistical mixture of coherent states,

$$\rho = \int P(\alpha_k) |\alpha_k\rangle\langle\alpha_k| d^2\alpha_k , \qquad (35)$$

where $P(\alpha_k)$ is a probability density, called $P$ representation, satisfying the normalization condition $\int P(\alpha_k) d^2\alpha_k = 1$ and $d^2\alpha_k = d(\text{Re}\,\alpha_k)d(\text{Im}\,\alpha_k)$. As shown by Glauber [16], if $\rho$ corresponds to a coherent state, $P(\alpha_k)$ is a Dirac δ-function. On the other hand, if $\rho$ represents a thermal Bose-Einstein distribution, $P(\alpha_k)$ will be a Gaussian function.

To conclude this section it is important to obtain the expectation values of the components of the magnetization operators for a single coherent state with eigenvalue $\alpha_k = |\alpha_k|\exp(i\phi_k)$. Using (29) in the expressions (27) it is straightforward to show that

$$\langle m_x(\vec{r},t)\rangle = \frac{M}{(NS/2)^{1/2}} |\alpha_k|(u_k + v_k)\cos(\vec{k}\cdot\vec{r} - \omega_k t + \phi_k) , \qquad (36a)$$

$$\langle m_y(\vec{r},t)\rangle = \frac{M}{(NS/2)^{1/2}} |\alpha_k|(u_k - v_k)\sin(\vec{k}\cdot\vec{r} - \omega_k t + \phi_k) . \qquad (36b)$$

The transverse components of the magnetization in (36) together with $\hat{z}M_z$ correspond to the classical view of a spin wave, namely, the magnetization precesses around the equilibrium direction with a phase that varies along the direction of propagation and with an ellipticity given by

$$\frac{m_x^{max}}{m_y^{max}} = \left(\frac{u_k + v_k}{u_k - v_k}\right) = \frac{A_k + B_k}{\omega_k} . \qquad (37)$$

Note that the elliptical precession of the transverse magnetization with frequency $\omega_k$ results in an oscillation of the $z$-component with frequency $2\omega_k$. As is well known it is this fact that makes possible to excite spin-waves with a microwave field parallel to the static field.

## IV- Microwave excitation of spin waves

Spin waves can be nonlinear excited in a magnetic material by means of several techniques employing microwave radiation, with the microwave magnetic field applied either perpendicular or parallel to the static field. The excitation is provided by the oscillation in the coupling parameter between two or more magnon modes, so the processes are called parametric. As in other nonlinear processes, the excitation occurs when the driving field exceeds a certain threshold value which depends on the rate at which the magnon mode relaxes to the heat bath. In the parallel pumping process the driving Hamiltonian in (8) follows from the Zeeman interaction of the microwave pumping field $\hat{z}h\cos(\omega_p t)$ with the magnetic system. One can express the Zeeman interaction in terms of the magnon operators using (9c), (11) and (21) and keeping only terms that conserve energy and show that the driving Hamiltonian for a ferromagnetic film is given by

$$H'(t) = \frac{\hbar}{2}\sum_k h\rho_k e^{-i\omega_p t} c_k^+ c_{-k}^+ + h.c. \quad , \tag{38a}$$

where

$$\rho_k = \gamma u_k v_k = \gamma \omega_M [(1-F_k)\sin^2\theta_k - F_k]/4\omega_k \tag{38b}$$

represents the coupling of the pumping field $h$ (frequency $\omega_p$) with the $\vec{k}, -\vec{k}$ magnon pair with frequency $\omega_k$ equal or close to $\omega_p/2$. Note that for a thick film, or a large wavevector, or a combination of both such that $kd >> 1$ Equation (20c) gives $F_k << 1$. In this case the coupling coefficient approaches the value for bulk samples $\rho_k = \gamma\omega_M \sin^2\theta_k/4\omega_k$. This is maximum for waves propagating perpendicularly to the field since they have the largest ellipticity and vanishes for waves with $\vec{k}$ along the field. However in films with $kd$ on the order of 1 or less $F_k$ is finite and the parallel pumping field can drive waves with any value of $\theta_k$. This is what happens in the case of the experiments in [1-6] with $H = 1.0$ kOe. As seen in Figure 1 (b) in a YIG film with $d = 5$ μm magnons with frequency 4.05 GHz and $\theta_k = 0$ can have two values for $k$, approximately 2 x $10^3$ cm$^{-1}$ and 5 x $10^5$ cm$^{-1}$. The first value corresponds to $kd \cong 1$ and $F_k \cong 0.6$ and the second to $kd \cong 250$ and $F_k \sim 0$. This means that magnons with frequency 4.05 GHz and $k \sim$ 2 x $10^3$ cm$^{-1}$ with $\theta_k = 0$ have a finite ellipticity and can be parallel-pumped. In fact, as can be seen in Figure 1 (b), for $H = 1.0$ kOe only waves with $\theta_k$ in the range from 0 to about 50° can be pumped at $f_p/2 = 4.05$ GHz. It turns out that as $\theta_k$ increases with fixed frequency the wavevector $k$ increases so $F_k$ decreases. In a film with thickness $d = 5$ μm this approximately compensates the increase in the $\sin^2\theta_k$ term so that the factor $u_k v_k$ which determines the parallel-pumping coupling remains about 0.2 in the whole range of $\theta_k$, 0 -50°.

The Heisenberg equation of motion for the operators $c_k$ and $c_k^+$ with the Hamiltonian $H = H_0 + H'(t)$ given by (22) and (38) can be easily solved assuming that the pumping field is applied at $t = 0$ to give the evolution of the expectation value of the number of magnons,

$$\langle n_k(t)\rangle = \langle n_k(0)\rangle e^{2\lambda_k t} \quad , \tag{39a}$$

where

$$\lambda_k = [(h\rho_k)^2 - \Delta\omega_k^2]^{1/2} - \eta_k , \tag{39b}$$

$\langle n_k(0)\rangle$ is assumed to be the thermal number of magnons, $\Delta\omega_k = \omega_k - \omega_p/2$ is the detuning from the frequency of maximum pumping strength and $\eta_k$ is the magnon relaxation rate which was introduced phenomenologically in the equations of motion.

Equations (39) express the well known effect of the parallel-pumping excitation. Magnon pairs with frequency $\omega_k$ equal or close to $\omega_p/2$ and wavevectors $\vec{k}, -\vec{k}$ determined by the dispersion relation are driven parametrically and their population grow exponentially when the field amplitude exceeds a critical value $h_c$, given by the condition $\lambda_k = 0$ in (39b),

$$h_c = (\eta_k^2 + \Delta\omega_k^2)^{1/2} / \rho_k \quad . \tag{40}$$

The large increase in the magnon population enhances the nonlinear interactions causing a reaction that limits its growth. Due to energy and momentum conservation the important mechanism in this process is the four-magnon interaction, which can be represented by a Hamiltonian of the form [32-36]

$$H^{(4)} = \hbar \sum_{k,k'} (\tfrac{1}{2} S_{kk'} c_k^+ c_{-k}^+ c_{k'} c_{-k'} + T_{kk'} c_k^+ c_{k'}^+ c_k c_{k'}) \quad , \tag{41}$$

where the interaction coefficients are determined mainly by the dipolar and exchange energies. For the $k$-values relevant in the experiments [1-6] the contribution from the exchange energy is negligible compared to the dipolar [33]. The four-magnon dipolar Hamiltonian can be obtained from (17) using for $\vec{m}$ and $\vec{h}_{dip}$ the first and second terms of the expansions in (10), following procedures similar to those in Sec. II and keeping only terms with two creation and two annihilation magnon operators. The result has several terms with coefficients containing the form factor $F_k$ in (20c) and products of the parameters $u_k$ and $v_k$ in (24), as given in Ref. [19]. It turns out that for the conditions of the experiments, $F_k \ll 1$, $u_k \sim 1$ and $v_k \ll 1$, so that the coefficients in (41) are given approximately by $S_{kk'} = 2T_{kk'} = 2\omega_M / NS$. Using the Hamiltonian (8) with (41) as the interaction term one can write the Heisenberg equations for the operators $c_k$ and $c_k^+$ from which several quantities of interest can be obtained. One of them is the correlation function $\sigma_k$ defined by [36],

$$\sigma_k = <c_k c_{-k}> = n_k e^{i\varphi_k} e^{-i2\omega_k t} \quad , \tag{42}$$

where $n_k$ is the magnon number operator and $\varphi_k$ the phase between the states of the pair. From the equation of motion for $\sigma_k$ it can be shown that for $h > h_c$, in steady-state [36-38]

$$\langle n_k \rangle_{ss} = \frac{[(h\rho_k)^2 - \eta_k^2]^{1/2} - |\Delta\omega_k|}{2V_{(4)}} \quad , \tag{43}$$

where

$$V_{(4)} = S_{kk} + 2T_{kk} = 4\omega_M / NS \quad . \tag{44}$$

It can also be shown that the phase $\varphi_k$ varies from $-\pi/2$ to $\pi$ as $h$ increases from $h_c$ to infinity. In the range of pumping power of the experiments [1-6] $\varphi_k \sim -\pi/2$. By using methods of quantum statistical mechanics and the probability density defined in (35) it has been demonstrated that the magnon pairs excited by parallel-pumping are in coherent magnon states but this is so only when the four-magnon interaction is taken into account [38].

Equation (43) shows that magnon pairs with frequency within a certain range around $\omega_p/2$ are pumped by the microwave field when its amplitude exceeds a critical value given by

$$h_c = \frac{(\eta_k^2 + \Delta\omega_k^2)^{1/2}}{\rho_k} \quad . \tag{45}$$

Note that the population of the parametric magnons is maximum for $\omega_k = \omega_p/2$ and for the allowed $\theta_k$ that maximizes $\rho_k$. The modes with $\omega_k = \omega_p/2$ are excited when the field amplitude $h$ is larger than a critical value $h_c = \eta_k/\rho_k$. In the reported experiments the minimum $h_c$ corresponds to a critical power $p_c$ in the range of 100 μW to 1 mW determined by the experimental geometry and the spin-lattice relaxation rate in YIG, $\eta_{SL} \sim 2 \times 10^6$ s$^{-1}$ [1-6]. However, when very short microwave pulses are used, much higher power levels are required to reduce the rise time and to build up large magnon populations. In this case it is the larger magnetic relaxation rate, $\eta_m \sim 25\eta_{SL} = 5 \times 10^7$ s$^{-1}$ in the experiments [4,5], that must be overcome by the driving. So one can define a critical field $h_{c1} = \eta_m / \rho_k = h_c \eta_m / \eta_{SL}$ for driving magnons with short pulses. Using the fact that the driving microwave power $p$ is proportional to $h^2$, we can write from (43) an expression for the steady-state number of parametric magnons with frequency $\omega_k = \omega_p / 2$ as a function of power,

$$\langle n_k \rangle_{ss} = \frac{[(p - p_{c1})/p_{c1}]^{1/2}}{2V_{(4)}/\eta_m} \qquad (46)$$

where $p_{c1} = p_c (\eta_m / \eta_{SL})^2$. Using numbers appropriate for the experiments [4-5], $p_c$ = 100 µW, $\eta_m$ = (1/20 ns) = 5 x $10^7$ s$^{-1}$, $p_{c1}$ = 0.0625 W, $V_{(4)} NS = 4\omega_M$ = 1.24 x $10^{11}$ s$^{-1}$, for a driving power $p$ = 4 W, Equation (46) gives for the normalized number of parametric magnons $\langle n_k \rangle_{ss} / NS$ = 1.6 x $10^{-3}$. The number of magnons pumped by the microwave field is actually larger than this because many modes with frequency in the vicinity of $\omega_p / 2$ are also driven. From (46) one can write an approximate equation for the total number of magnons pumped into the system as

$$n_p = r_p n_H [(p - p_{c1})/p_{c1}]^{1/2} \qquad (47a)$$

where

$$n_H \equiv \eta_m / 2V_{(4)} = \eta_m NS / 8\omega_M \qquad (47b)$$

and $r_p$ is a factor that represents the number of pumped modes weighted by a factor relative to the number of magnons of the mode with maximum coupling.

## V. Model for Bose-Einstein condensation in the microwave driven interacting magnons

In the experiments of [1-6] magnon pairs are parametrically driven by parallel-pumping in a YIG film at large numbers compared to the thermal values. The population of these primary magnons with frequency equal or close to $\omega_p / 2$ is quickly redistributed over a broad frequency range down to the minimum frequency $f_{min} = \omega_{k0} / 2\pi$. This redistribution is caused by four-magnon scattering events which conserve the total number of magnons so that a quasi-equilibrium hot magnon gas is formed. Since the spin-lattice relaxation time in YIG is much longer than the intermagnon decay time, the hot magnon gas remains practically decoupled from the lattice for several hundred ns with an essentially constant number of magnons. In this situation the occupation number of the system is given by the Bose-Einstein distribution

$$n_{BE}(\omega, \mu, T) = \frac{1}{e^{(\hbar\omega - \mu)/k_B T} - 1} \qquad (48)$$

where $\mu$ is the associated chemical potential. As is well known [7] in systems with constant number of particles it is (48) and not (32) that determines the distribution of the number of bosons with energy $\hbar\omega$ at a given temperature $T$, provided the system is in equilibrium and there is no interaction between the bosons. The experiments of [1-5] were done with 8.1 GHz microwave pumping in two types of pulsed regimes and the properties of the pumped magnon system were measured with time-resolved Brillouin light scattering. In the first one long pulses of duration 1 µs were employed to ensure that quasi-equilibrium was established in the hot magnon gas while still decoupled from the lattice. This made possible the observation of the full thermal equilibrium spectra between $f_{min}$ and the parametric magnon frequency of 4.05 GHz as a function of the microwave pumping power. The authors of [1-5] argue that without external driving the magnons are in thermal equilibrium with the lattice and have uncertain number so that $\mu = 0$. If a microwave driving is applied and the power exceeds the threshold for parallel pumping the total number of particles in the magnon gas increases and can be expressed as

$$N_{tot} = \int D(\omega) n_{BE}(\omega, \mu, T) d\omega \tag{49}$$

where $D(\omega)$ is the magnon density of states and the integral in (49) is carried out over the whole range of magnon frequencies. Clearly as the microwave power is raised the total number of magnons increases so that the temperature and the chemical potential increase. Using (49) and the similar equation for the energy of the system it is possible to determine the values of $\mu$ and $T$ for a given $N_{tot}$. In the experiments with long pulses [1-3] the BLS spectra could be fitted with the spectral density function $D(\omega) n_{BE}(\omega, \mu, T)$, allowing the determination of $\mu$ and $T$ for each power value. At a high enough power the chemical potential reaches the energy corresponding to $f_{min}$ resulting in an overpopulation of magnons with that frequency relative to the theoretical fit. It was then necessary to add a singularity at $f_{min}$ to fit the spectrum [2]. This was interpreted as a signature of the Bose-Einstein condensation of magnons, namely: when the number of magnons reaches a critical value defined by the condition $\mu_c = \hbar 2\pi f_{min}$ the gas is spontaneously divided in two parts, one with the magnons distributed according to (48) and another one with the magnons accumulated in the state of minimum energy.

The experiments with short microwave pulses (30 ns) [4,5] allowed the observation of the dynamics of the redistribution of energy from the primary magnons to the modes in the broader energy range and the formation of the strong BLS peak at $f_{min}$. The behavior of the peak intensity and of the relaxation to the lattice with increasing microwave pumping power revealed that above a critical power level the magnons accumulated at the bottom of the spectrum develop a spontaneous emergence of coherence. The coherence of the BEC was further confirmed in experiments showing the microwave emission from the $k = 0$ mode generated by the coalition of a pair of BEC magnons when the applied field has a value for which its frequency is $2 f_{min}$ [6]. While the thermodynamic interpretation of the experiments in [1-5] is quite satisfactory and explains qualitatively several observed features, it fails in providing quantitative results to compare with data and, most serious, it does not explain the observed spontaneous emergence of quantum coherence in the BEC of magnons. This is not surprising because a system of free noninteracting magnons cannot possibly evolve spontaneously from quantum states describing thermal magnons, represented by the distribution (33), to coherent magnons states corresponding to (31). The theory presented in this section shows that the cooperative action of the magnon gas through the four-magnon interaction can provide the mechanism for the observed spontaneous emergence of quantum coherence in the BEC. The theory relies in part on some assumptions based on the experimental observations and on some approximations to allow an analytical treatment of the problem. The ultimate justification for the assumptions and approximations is the good agreement of the theoretical results with the experimental data for the BLS intensity and for the emitted microwave signal as a function of the microwave pumping power presented in the next section.

We consider that with microwave pumping the magnon system can then be decomposed in two sub-systems, one with frequency above $\omega_p/2$ in thermal equilibrium with the lattice at room temperature and another one with frequency in the range $\omega_{k0} - \omega_p/2$ in quasi-equilibrium at a higher temperature $T$. The second sub-system, which we call the magnon reservoir, is characterized by an occupation number given by the Bose-Einstein distribution with its own temperature and chemical potential. We also assume that after the hot magnon reservoir is formed by the redistribution of the primary magnons, the correlation between the phases of the magnon pairs lasts for a time that can be as large as $4/\eta_m$, which is about 100 ns in the experiments [1-6]. This is a sufficient time for the four-magnon interaction to come into play for establishing a cooperative phenomenon to drive a specific $k$ mode. The effective driving Hamiltonian for this process is obtained from Equation (41) by taking averages of pairs of destruction operators for reservoir magnons to form correlation functions as defined in (42),

$$H'(t) = \hbar \sum_{k_R} \tfrac{1}{2} S_{kk_R} n_{k_R} e^{i\varphi_{k_R}} e^{-i2\omega_{k_R} t} c_k^+ c_{-k}^+ + h.c. \quad . \tag{50}$$

Equation (50) has a form that resembles the Hamiltonian (38) for parallel pumping, revealing that under appropriate conditions magnon pairs can be pumped out of equilibrium in the gas. To treat (50) we note that since the number of the magnons $n_p$ pumped into the system is much larger than the number of thermal magnons in the range $\omega_{k0} - \omega_p/2$ one can write for the magnon reservoir

$$n_p = \int D(\omega) n_{BE}(\omega, \mu, T) d\omega \quad , \tag{51}$$

where $n_p$ is related to the power as in (47). Of course the calculation of the population in each state of the reservoir as a function of power is a formidable task. So we use some approximations to treat the problem analytically. Consider that the population of the primary magnons is distributed among the $N_R$ modes $k_R$ in the magnon reservoir, so that with (47) we can write an expression for the average population of modes $k_R$ as a function of pumping power $p$,

$$n_R = r n_H [(p - p_{c1})/p_{c1}]^{1/2} \quad , \tag{52a}$$

where

$$r = r_p / N_R \quad . \tag{52b}$$

If all the reservoir states had the same magnon number the sum in $k_R$ in (50) would reproduce the density of states $D(\omega)$. Actually the number of magnons in each state $k_R$ depends on its energy as given by (48) and can be written approximately as $n_{k_R} = f_{BE}(\omega_{kR}) n_R$, where $f_{BE}(\omega_{kR})$ is a function proportional to (48) with a normalization constant so that its average over the frequency range of the reservoir modes is unity,

$$f_{BE}(\omega) = n_{BE}(\omega)/C_{BE} \quad , \tag{53a}$$

$$C_{BE} = \frac{1}{\Delta \omega_R} \int n_{BE} d\omega \quad , \tag{53b}$$

$\Delta \omega_R = \omega_p/2 - \omega_{k0}$ being the frequency range of the reservoir modes. Thus the relevant quantity for determining the frequency dependence of the coefficient in the Hamiltonian (50) is the density of states weighted by the normalized Bose-Einstein distribution,

$$G(\omega) = D(\omega) f_{BE}(\omega) \quad . \tag{54}$$

Note that $f_{BE}(\omega)$ and $G(\omega)$ also vary with $\mu$ and $T$ but we omit them in the functions to simplify the notation. Figure 3 shows plots of (54) for several values of $\mu$ and the corresponding $T$ for a 5 μm thick YIG film. The density of states was calculated numerically using the approximate dispersion relation (26) by counting the number of states with $k_x = \pm n_x 2\pi/L_x, k_z = \pm n_z 2\pi/L_z$ having frequencies in discrete intervals $\delta \omega = 2\pi \times 1.0$ MHz in the range $0 - \omega_p/2$. The value of $\mu$ were chosen so that their differences to $\hbar \omega_{k0}$ are the same as the ones used in [3] to fit the measured BLS spectra with varying microwave power. The corresponding values of $T$ were estimated by the fits to the BLS spectra in [3]. The dimensions used to calculate the density of states are $L_x = L_z = 2$ mm. As expected $G(\omega)$ has a peak at the minimum frequency that becomes sharper the chemical potential rises and approaches the minimum energy. The consequence of this is that as the microwave pumping power increases and $(\hbar \omega_{k0} - \mu)/k_B T$ becomes very small the peak in $G(\omega)$ dominates the coefficient in (50) revealing that it is possible to establish a cooperative action of the modes with frequency $\omega_{k_R}$ close to $\omega_{k0}$ so as to drive magnon pairs nonlinearly as in the parallel pumping process. Considering that the pumping is effective for frequencies $\omega_{k_R}$ in the range $\omega_{k0} \pm \eta_m$, the sum over $k_R$ in (50) can be replaced by $D(\omega_{k0}) \eta_m$ so that one can write an effective Hamiltonian for driving $\vec{k}_0, -\vec{k}_0$ magnon pairs as,

$$H'_{eff}(t) \cong \hbar (h\rho)_{eff} e^{-i 2\omega_{k0} t} c^+_{k_0} c^+_{-k_0} + h.c. \quad , \tag{55a}$$

where

$$(h\rho)_{eff} = -i G(\omega_{k0}) \eta_m V_{(4)} n_R / 2 \tag{55b}$$

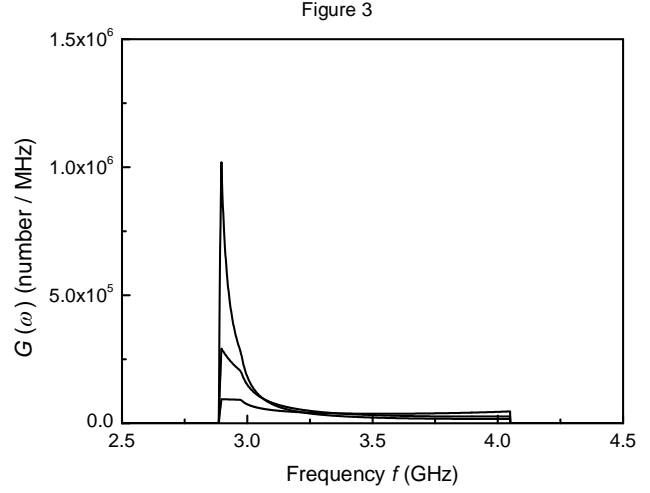

Figure 3: $G(\omega)$ as a function of frequency for spin waves in a 5 μm thick YIG film in a field $H = 1.0$ kOe with the following parameters: $\mu = 0$, $T = 300$ K (lowest values at $f_{min} = 2.898$ GHz); $\mu / h = 2.718$ GHz, $T = 900$ K; $\mu / h = 2.868$ GHz, $T = 1200$ K ( h is Plank's constant).

represents an effective field proportional to the average number of magnons $n_R$ in the reservoir. Note that the factor $-i$ in (55b) arises from the phase between pairs that is approximately $-\pi/2$ in the range of power of interest. From the analysis in Sec. IV one can see that there is a critical number of reservoir modes above which they act cooperatively to pump the $\vec{k}_0, -\vec{k}_0$ magnons parametrically. The condition $|(h\rho)_{eff}| = \eta_m$ gives the critical average number of reservoir magnons

$$n_c = 2/V_{(4)} G(\omega_{k0}) \qquad . \qquad (56)$$

Since the Hamiltonian (55) has the same form as (38), the population of the $k_0$ mode driven by the effective field and saturated by the effect of the four-magnon interaction is calculated in the same manner as done for the direct parallel-pumping process. Thus from (43) with $\Delta\omega_k = 0$ we have

$$n_{k0} = \frac{[|(h\rho)_{eff}|^2 - \eta_m^2]^{1/2}}{2V_{(4)}} \qquad . \qquad (57)$$

Using (47b), (55b) and (56) in Equation (57) one can write the population of the $k_0$ mode in terms of the average reservoir number $n_R$,

$$n_{k0} = \frac{n_H}{n_c}(n_R^2 - n_c^2)^{1/2} \qquad . \qquad (58)$$

Alternatively $n_{k0}$ can be written in terms of the pumping power using (52) and (56) in (58),

$$n_{k0} = n_H [(p - p_{c2})/(p_{c2} - p_{c1})]^{1/2} , \qquad (59)$$

where $n_H$ is given by (47b) and

$$p_{c2} = p_{c1}\{1 + 16/[r\eta_m G(\omega_{k0})]^2\} \qquad (60)$$

is another threshold power level $p_{c2} \gg p_{c1}$. Note that with (52) and (60) the effective driving field (55b) can be expressed in terms of power as

$$(h\rho)_{eff} = -i\eta_m [(p - p_{c2})/(p_{c2} - p_{c1})]^{1/2} \qquad . \qquad (61)$$

Notice that since $G(\omega_{k0})$ depends on $\mu$ and consequently on the power, the value of $\mu$ that enters in (56) and (60) is the one for $p = p_{c2}$. Equations (58) and (59) are valid only for $n_R \geq n_c$ or equivalently $p \geq p_{c_2}$ and they represent the first important result of this paper. For $n_R < n_c$, or $p < p_{c2}$ the population of the $k_0$ mode is that of thermal equilibrium with the reservoir given by

$$\bar{n}_{k0} = n_R f_{BE}(\omega_{k0}) \ . \tag{62}$$

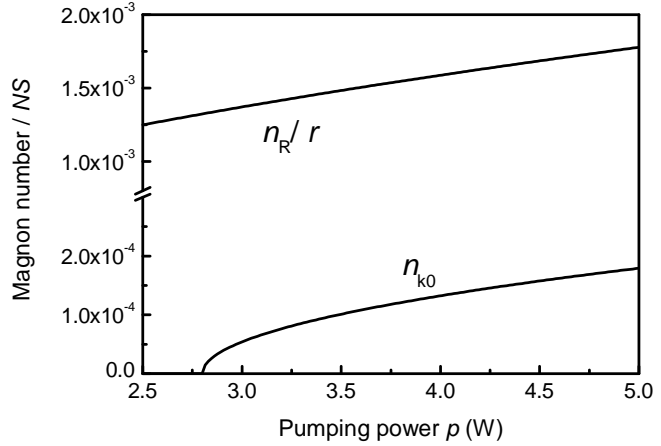

Figure 4: Variation with microwave pumping power of the normalized reservoir average magnon number and of the BEC magnon population.

However, for $n_R \geq n_c$ or $p \geq p_{c_2}$ the population of mode $k_0$ is pumped-up out of equilibrium as a result of a spontaneous cooperative action of the reservoir modes. As it will be shown in the next section the $k_0$ mode with population given by (57)-(59) above the threshold is in a coherent magnon state. This means that when the average reservoir magnon number reaches the critical value (56) the magnon gas separates in two parts, one in thermal equilibrium with the reservoir having frequencies in a wide range and one with a higher magnon number in a narrow range around the minimum frequency. This is one of the characteristic features of a Bose-Einstein condensate.

We now have the necessary elements to interpret the behavior of the magnon system with increasing microwave pumping power. First we note that in the interacting magnon gas the formation of the BEC occurs at a value of the chemical potential that is close but not equal to the minimum energy $\hbar\omega_{k0}$. This is so because as the microwave power increases and $\mu$ approaches $\hbar\omega_{k0}$, the average reservoir number reaches the critical value (56) corresponding to a small but finite $(\hbar\omega_{k0} - \mu)$. The value of the chemical potential satisfying (56) can be identified as the critical value $\mu_c$ for the formation of the BEC. Using (48), (53), (54) and (60) one can obtain the following relation between $\mu_c$ and the critical power $p_{c2}$

$$\hbar\omega_{k0} - \mu_c = \frac{r\eta_m D(\omega_{k0}) k_B T}{4 C_{BE}} [(p_{c2} - p_{c1})/p_{c1}]^{1/2} \quad , \tag{63}$$

where we have considered that $(\hbar\omega_{k0} - \mu_c)/k_B T \ll 1$ to use the binomial expansion of the exponential function in (48). Of course Equation (63) is not an explicit expression for the critical chemical potential in terms of $p_{c2}$ because $C_{BE}$ and also the effective temperature $T$ vary with $\mu$. Equation (63) is important to demonstrate that the difference $(\hbar\omega_{k0} - \mu_c)$ is finite in the interacting magnon gas. As the microwave power increases above $p_{c1}$ the average reservoir magnon number $n_R$ increases continuously as given by (52). The variation of $n_R$ with $p$ is shown in Figure 4. Correspondingly the chemical potential increases with power and reaches the critical value $\mu_c$ when $p$ reaches $p_{c2}$, giving rise to the nonlinear driving of the $k_0$ mode. This process leads to a sharp increase in the magnon population at the state with minimum frequency $\omega_{k0}$ characteristic of the condensation of bosons. Thus the population $n_{k0}$ will henceforth be called condensate or BEC magnon number. For $p \geq p_{c_2}$ the chemical potential locks at the value $\mu_c$ so that the dependence of $(h\rho)_{eff}$ on power is entirely contained in (61). Since the four-magnon interaction that produces the cooperative action conserves the number of magnons, as $p$ increases further the number of magnons in the reservoir stays constant and the additional magnons originating from the primary magnons end up at the condensate state. Figure 4 also shows the variation with power of the BEC number $n_{k_0}$ for $p \geq p_{c2}$.

## VI. Quantum coherence of the Bose-Einstein condensate

In order to study the coherence properties of the $k_0$ mode pumped above threshold one has to use methods of statistical mechanics appropriate for boson systems interacting with a heat-bath. We follow here the same procedure used to study the direct parallel pumping process [38]. The first step is to represent the magnon reservoir and its interactions with a specific $k$ mode by a Hamiltonian that allows a full description of the thermal and driving processes for the interacting magnon system,

$$H = H_0 + H^{(4)} + H'_{\text{eff}}(t) + H_R + H_{RS}, \tag{64}$$

where the first three terms are given respectively by (22), (41) and (55),

$$H_R = \hbar \sum_{k_R} \omega_{kR} c^+_{kR} c_{kR} \tag{65}$$

is the Hamiltonian for the magnon reservoir, assumed to be a system with large thermal capacity and in thermal equilibrium and

$$H_{RS} = \hbar \sum_{k,k_R} \beta^*_{k,kR} c^+_{kR} c_k + \beta_{k,kR} c_{kR} c^+_k \tag{66}$$

represents a linear interaction between the magnons $k$ and the heat reservoir. Note that (66) also has its origin in the four-magnon interaction which provides the main mechanism for the intermagnon relaxation. Using the Heisenberg equation for the magnon operators for a mode $k$ in the vicinity of $k_0$ with the total Hamiltonian (64) and assuming that $n_k = n_{-k}$ we obtain,

$$\frac{dc_k}{dt} = -(i\omega_k + \eta_m + i2V_{(4)} n_k) c_k - i(h\rho)_{\text{eff}} e^{-i2\omega_{k0}t} c^+_{-k} + F_k(t) \tag{67}$$

where

$$\eta_m = \pi D(\omega_k) |\beta_{k,kR}|^2 , \tag{68a}$$

$$F_k(t) = -i \sum_{k_R} \beta_{k,kR} c_{kR} e^{-i\omega_{kR} t} , \tag{68b}$$

represent respectively the magnetic relaxation rate expressed in terms of the interaction between magnon $k$ and the heat reservoir and a Langevin random force with correlators of Markoffian systems type [38-40]. Using Equation (67) and the corresponding one for the operator $c^+_{-k}$, transforming them to the representation of coherent magnon states $|\alpha_k\rangle$ and working with variables in a rotating frame $c_k |\alpha_k\rangle = \alpha_k(t) e^{-i\omega_k t} |\alpha_k\rangle$ we obtain an equation of motion for coherent state eigenvalue with $k = k_0$,

$$\frac{d\alpha_k(t)}{dt} - \frac{2V^2_{(4)}}{\eta_m} \left( \frac{|(h\rho)_{\text{eff}}|^2 - \eta^2_m}{4V^2_{(4)}} - |\alpha_k(t)|^4 \right) \alpha_k(t) = S_k(t) \tag{69a}$$

where

$$S_k(t) = F_k(t) e^{i\omega_k t} - i \frac{(h\rho)_{\text{eff}}}{\eta_m} F^*_{-k}(t) e^{-i\omega_k t} . \tag{69b}$$

Equations (69) contain all the information carried by the equations of motion for the magnon operators. It is a typical nonlinear Langevin equation which appears in Brownian motion studies and laser theory [39,40]. It shows that the magnon modes with amplitude $\alpha_k$ are driven thermally by the hot magnon reservoir and also by an effective driving field. The solutions of (69) confirm the previous analysis. For negative values of the driving term $[|(h\rho)_{eff}|^2 - \eta_m^2]$ the magnon amplitudes are essentially the ones of the thermal reservoir. For positive values they grow exponentially and are limited by the effect of the four-magnon interactions. Above the threshold condition the steady-state solution of (69) gives for the number of magnons $n_k = |\alpha_k|^2$ an expression identical to (57). The final step to obtain information about the coherence of the excited mode is to find an equation for the probability density $P(\alpha_k)$, defined in (35), that is stochastically equivalent to the Langevin equation. Using $\alpha_k = a_k \exp(i\phi_k)$ we obtain a Fokker-Plank equation in the form [38],

$$\frac{\partial P}{\partial t'} + \frac{1}{x}(\frac{\partial}{\partial x})[(A - x^4)x^2 P] = \frac{1}{x}\frac{\partial}{\partial x}(x\frac{\partial P}{\partial x}) + \frac{1}{x^2}(\frac{\partial^2 P}{\partial \phi_k^2}) \quad , \tag{70}$$

where

$$t' = (\bar{n}_{k0}^2 \eta_m^3 / n_H^2)^{1/3} t \quad , \tag{71a}$$

$$x = (2/n_H^2 \bar{n}_{k0})^{1/6} a_k \tag{71b}$$

represent normalized time and magnon amplitude and the parameter $A$ is given by,

$$A = (\frac{2}{n_H^2 \bar{n}_{k0}})^{2/3} \frac{[|(h\rho)_{eff}|^2 - \eta_m^2]^{1/2}}{2V_{(4)}} \equiv (\frac{2}{n_H^2 \bar{n}_{k0}})^{2/3} n_{k0}^2 \quad . \tag{71c}$$

Note that $A$ can alternatively be written in terms of the average reservoir number $n_R$ or the power $p$ as,

$$A = [\frac{2n_H}{f_{BE}(\omega_{k0})n_R}]^{2/3} [(n_R/n_c)^2 - 1] \quad , \tag{72a}$$

$$A = [\frac{2}{rf_{BE}(\omega_{k0})}]^{2/3} [\frac{p_{c1}}{p - p_{c1}}]^{1/3} \frac{(p - p_{c2})}{(p_{c2} - p_{c1})} \quad . \tag{72b}$$

Application of Equation (70) to describe the full dynamics of the pulsed experiments [1-6] must consider that the factors relating $t'$ to $t$ and $x$ to $a_k$, as well as the parameter $A$, are all time dependent. However, for typical numbers appropriate for the experiments, $t' \sim t \times 2\times 10^6$ s$^{-1}$, so that the dynamics of the pulsed experiments is relatively slow in the renormalized time scale. Thus in a first approximation we assume that all parameters are constant and obtain the stationary solution of (70) independent of $\phi_k$ in the form,

$$P(x) = C\exp(\tfrac{1}{2}Ax^2 - \tfrac{1}{6}x^6) . \tag{73}$$

where $C$ is a normalization constant such that the integral of $P(x)$ in the range of $x$ from zero to infinity is equal to unity. Note that for obtaining (73) all integration constants were set to zero to satisfy this condition. Figure 5 shows plots of $P(x)$ for four values of the parameter $A$, -1, 0, 80 and 250. In choosing the positive values we have considered parameters which enter in (72a) and (72b) appropriate for the experiments [1-6]: $p_c = 100$ mW, $p_{c1} = 0.0625$ W and $p_{c2} = 2.8$ W; $rf_{BE}(\omega_{k0}) \sim 8 \times 10^{-7}$ obtained from the fit of theory to the BLS data as shown in the next Section. With these numbers we obtain $A = 250$ for $n_R/n_c = 1.023$, or equivalently $p/p_{c2} = 1.047$.

Equation (72a) shows that for reservoir average populations below the critical number, $n_R < n_c$, the parameter $A$ is negative. In this case the function $P(x)$ in (73) behaves as a Gaussian distribution, characteristic of systems in thermal equilibrium and described by incoherent magnon states [16]. On the other hand for $n_R > n_c$, or $p > p_{c2}$, $A > 0$ and the stationary state consists of two components, a coherent one convoluted with a much smaller fluctuation with Gaussian distribution. Since the variance of $P(x)$ is proportional do $A^{-1}$, for $A \gg 1$ the function $P(x)$ becomes a delta-like distribution, characteristic of a coherent magnon state [16]. Figure 5 shows that in the conditions of the experiments $P(x)$ becomes a delta-like function at power levels just above the critical value. Note that only in the presence of the four-magnon interaction the magnon state driven collectively by the reservoir modes is a coherent state [38]. Note also that $P(x)$ has a peak at $x_0 = A^{1/4}$, so that it represents a coherent state with an average number of magnons given by $x_0^2 = A^{1/2}$. From (71b) and (71c) we see that this corresponds to a magnon number $a_0^2$ which is precisely the value $n_{k_0}$ given by (58) and (59). This means that the magnon $\omega_{k_0}$ driven cooperatively by the reservoir modes is a quantum coherent state. This is the second and most important result of this paper since the coherence implies a macroscopic wavefunction satisfying an essential condition for the condensate.

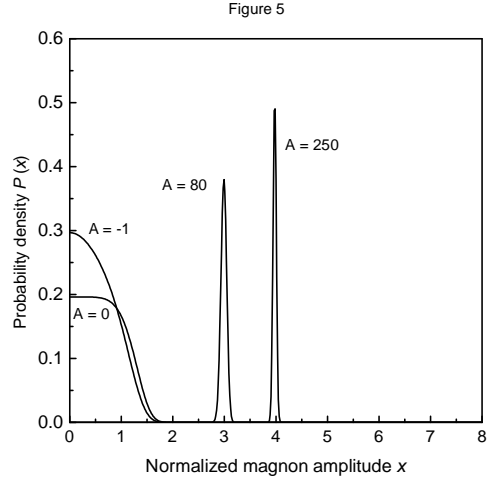

Figure 5: Probability density characteristic of a microwave driven interacting magnon system for several values of the parameter $A$: Negative values correspond to $n_R < n_c$ or for $p < p_{c2}$; $A = 0$ corresponds to the threshold; $A = 80$ and $250$ correspond to $n_R/n_c = 1.008$ and $1.023$, or to $p/p_{c2} = 1.015$ and $1.047$.

The theoretical interpretation of the observations of Demokritov and co-workers [1-6] is now clear. After the reservoir of hot magnons with population $n_R$ is formed as a result of the fast redistribution of the energy of the primary parallel-pumped magnons, the modes with frequency $\omega_{k_R}$ close to $\omega_{k_0}$ act together to drive the mode $k_0$. However only if the microwave power is above a critical value $p_{c2}$, $n_R$ exceeds $n_c$ and the system spontaneously develops a coherent state with frequency $\omega_{k_0}$. According to (36) the small-signal dynamic magnetization is proportional to the amplitude of the coherent state, $m^+ \propto a_{k_0} = n_{k_0}^{1/2}$. Thus one can write from (59) that for $p > p_{c2}$ the dynamic magnetization scales with microwave power as $m^+ \propto (p - p_{c2})^{1/4}$, characteristic of a second-order phase transition. The spontaneous emergence of quantum coherence [41] caused by a phase-transition and the associated magnetic dynamic order in a macroscopic scale, constitute rigorous theoretical support for the formation of Bose-Einstein condensation of magnons at room temperature, as claimed by Demokritov and co-workers [1-6].

## VII. Comparison with experimental data

In this section we apply the model for the formation of the BEC of magnons developed to treat an interacting magnon gas driven by microwave radiation in a YIG film. We compare the results of the theory with the data obtained by Demokritov and co-workers [1-6] using two very different techniques, Brillouin light scattering form the magnon condensate and microwave emission from the uniform mode driven by BEC magnon pairs. In both cases the theory developed here allows the calculation of quantities of interest as a function of microwave pumping power to compare with data.

## a- Intensity of the Brillouin Light Scattering

In the experiments of [4,5] with short microwave pulse driving the coherence properties of the excited magnons states emerge clearly in the behavior of the intensity of the BLS peak at $f_{min}$. As argued in [4,5], for incoherent scatterers the BLS intensity is proportional to their number, whereas for coherent scatterers it is proportional to the number squared. Thus, in order to compare theory with data we express the BLS intensity in terms of the microwave power $p$ in two regimes: for $p < p_{c2}$ the number of magnons with frequency $f_{min}$ is the thermal number $\bar{n}_{k0}$ given by (52) and (62); for $p \geq p_{c2}$ the condensate is characterized by a coherent magnon state with number given by (59). Consider that the relevant number of scatterers is the number of magnons per spin site. Using (52) and (62) we obtain for $p < p_{c2}$,

$$I^{inc} = b \left( \frac{\bar{n}_{k0}}{N} \right) = b \, r \, f_{BE}(\omega_{k0}) \left( \frac{n_H}{N} \right) \left( \frac{p - p_{c1}}{p_{c1}} \right)^{1/2}, \tag{74}$$

and with (59) we have for $p > p_{c2}$,

$$I^{coh} = b \left( \frac{n_{k0}}{N} \right)^2 = b \left( \frac{n_H}{N} \right)^2 \left( \frac{p - p_{c2}}{p_{c2} - p_{c1}} \right), \tag{75}$$

where $b$ is a scale factor proportional to the magneto-optical constant and involves electromagnetic, magnetic and geometrical quantities. Figure 6 shows a fit of (74) and (75) to data, using $I^{inc} = c_1 (p - p_{c1})^{1/2}$ and $I^{coh} = c_2 (p - p_{c2})$, with $c_1 = 6.7$, $c_2 = 370.0$ and $p_{c2} = 2.8$ W. Using (47b), (74) and (75) one can obtain a relation that allows the calculation of the factor $r f_{BE}(\omega_{k0})$ at the critical chemical potential from the fitting parameters,

$$r f_{BE}(\omega_{k0}) = \frac{c_1}{c_2} \frac{\eta_m}{8 \omega_M} \frac{p_{c1}^{1/2}}{p_{c2}}, \tag{76}$$

from which we obtain $r f_{BE}(\omega_{k0}) = 8.1 \times 10^{-7}$. Before discussing the implications of this result it is interesting to compare its value with the one obtained directly from the measured $p_{c2} = 2.8$ W. Using this value and $p_{c1} = 0.0625$ W in (52) and (60) we obtain $r f_{BE}(\omega_{k0}) D(\omega_{k0}) \eta_m = 0.6$. Considering $D(\omega_{k0}) \approx 10^5 /$ MHz, calculated numerically as described earlier, and $\eta_m / 2\pi = 8$ MHz, we find $r f_{BE}(\omega_{k0}) = 7.4 \times 10^{-7}$, which is very close to the value obtained from (76).

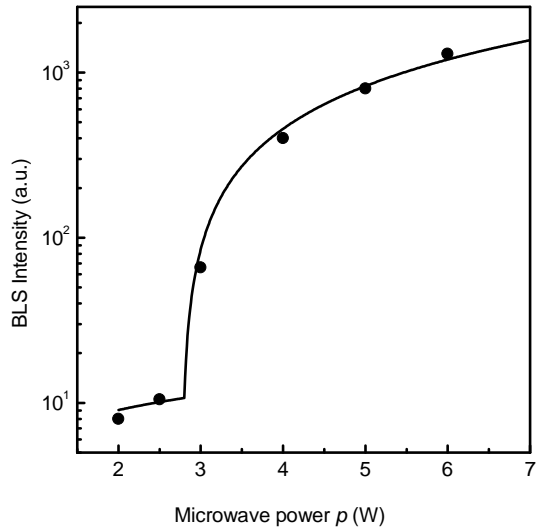

Figure 6: Fit of the theoretical result (solid line) for the BLS intensity as a function of microwave pumping power to the experimental data (symbols) of Demokritov and co-workers [4,5].

To obtain a value for $f_{BE}(\omega_{k0})$ at the critical chemical potential we use the definition (53a) and consider that the difference between the minimum energy $\hbar\omega_{k0}$ and $\mu_c$ is, in frequency units, in the range (10 – 20) MHz. The normalization constant $C_{BE}$ is calculated by the integration of (53b) in the frequency range (2.9 – 4.05) GHz using the binomial expansion of the exponential in (48) and assuming $T = 10^3$ K. We obtain $C_{BE} = (0.8 - 0.9) \times 10^5$ and $f_{BE}(\omega_{k0}) \approx (10 - 25)$ for the range of $\mu_c$ above. With these values we have an order of magnitude estimate for $r = r_p/N_R \sim 10^{-7}$. Considering the number of reservoir states $N_R \sim 10^8 - 10^9$ obtained numerically we find for the pumping factor $r_p \sim 10 - 10^2$. This is quite small compared to the value $r_p \sim 10^3 - 10^4$ calculated numerically by counting the states with frequency in the range $\omega_p/2 \pm \eta_m$ for the conditions of the experiments. We attribute this discrepancy to one or a combination of the following reasons: a flaw in the theoretical model; a failure of Equation (26) in reproducing the correct slopes of the dispersion curves near the frequency minima introducing considerable error in the calculation of density of states; a large number of magnons is lost on the way to the region of minimum frequency in the process of redistribution of the primary magnon population.

**b- Microwave emission from the BEC of magnons**

As observed by Dzyapko *et al*. [6], if the static field applied to a microwave pumped YIG film has a value such that the frequency of the $k \approx 0$ magnon is $\omega_0 = 2\omega_{k0}$, a microwave signal is emitted with frequency $\omega_0$. They interpret this radiation as due to $k \approx 0$ magnons created by pairs of BEC magnons $\vec{k}_0, -\vec{k}_0$ through a three-magnon confluent process. The $k \approx 0$ value is necessary for emission because the wavenumber of electromagnetic radiation with frequency 1.5 GHz, as in the experiments [6], is $k = 2\pi/f \approx 0.3$ cm$^{-1}$. Figure 7 illustrates the three-magnon confluent process in the dispersion relation for modes propagating along the field in a 5 µm thick YIG film for $H = 520$ Oe, which is the field value for $\omega_0 = 2\omega_{k0}$. As we have shown earlier [20], the $k \approx 0$ magnons created by the BEC are coherent magnons states. Thus they correspond to a nearly uniform magnetization precessing with frequency $\omega_0$ that emits electromagnetic radiation with this frequency [42-44]. To calculate the power emitted by the uniform mode as a function of the microwave pumping power we need to study the process by which this mode is driven by the BEC magnon pairs. Consider a Hamiltonian as in (8) in which the magnon interaction include three- and four-magnon contributions,

$$H = H_0 + H^{(3)} + H^{(4)} + H'_{eff}(t) \quad , \tag{77}$$

where $H'_{eff}(t)$ is the effective Hamiltonian for driving $\vec{k}_0, -\vec{k}_0$ magnon pairs given by (55) and (61) and the Hamiltonian for the three-magnon confluence process is [32-34]

$$H^{(3)} = \hbar V_{(3)} c_0^+ c_{k0} c_{-k0} + h.c. \quad , \tag{78}$$

where the vertex of the interaction for small wavevectors is dominated by the dipolar interaction between the spins $S$ and is given approximately by $V_{(3)} = \omega_M/(2SN)^{1/2}$. To study the process by which the pairs of BEC coherent magnons $\vec{k}_0, -\vec{k}_0$ are produced and then generate $k \sim 0$ modes we use the Hamiltonian (77) to obtain the Heisenberg equations of motion for the magnon operators,

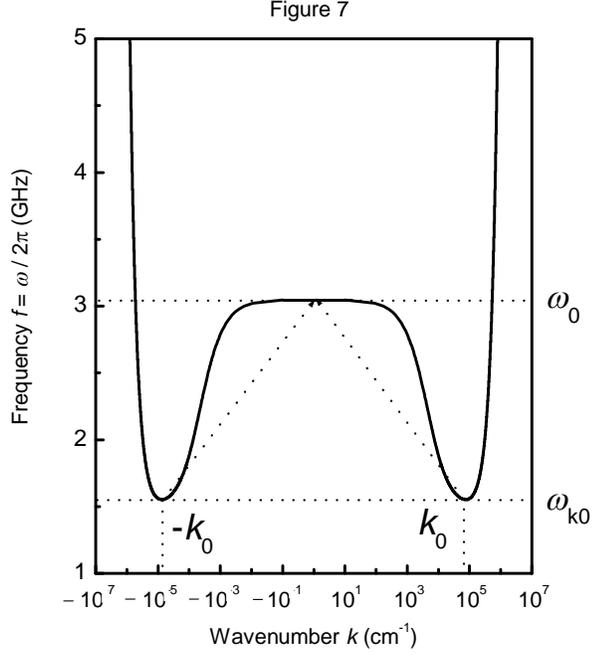

Figure 7: Dispersion relation for magnons propagating along the field $H = 520$ Oe applied in the plane of a YIG film with thickness 5 μm with illustration of the 3-magnon coalescence process that generates a $k = 0$ from a pair of BEC magnons.

$$\frac{dc_0}{dt} = -(i\omega_0 + \eta_0 + iV_{(4)} n_0) c_0 - iV_{(3)} c_{k0} c_{-k0} \quad , \qquad (79)$$

$$\frac{dc_{k0}}{dt} = -(i\omega_{k0} + \eta_{k0} + i2V_{(4)} n_{k0}) c_{k0} - i[V_{(3)} c_0 + (h\rho)_{\mathit{eff}}\, e^{-i2\omega_{k0} t}]c^+_{-k0}, \qquad (80)$$

where the relaxation was introduced phenomenologically. We consider that all states involved are coherent magnon states as demonstrated earlier and work with the corresponding eigenvalues $\alpha_k$. In addition we assume that there are $p_{k0}$ pair modes with wavevectors close to $\vec{k}_0, -\vec{k}_0$ to drive the $k \sim 0$ modes and that the resonance condition is satisfied $\omega_0 = 2\omega_{k0}$, determined by the value of the applied field $H$. The equations of motion for the eigenvalue $\alpha_0$ and the correlation function $\sigma_{k0} = \alpha_{k0} \alpha_{-k0}$ in a frame rotating with frequency $\omega_0$ become,

$$\frac{d\alpha_0}{dt} = -(\eta_0 + iV_{(4)} n_0)\alpha_0 - ip_{k0} V_{(3)} \sigma_{k0} \quad , \qquad (81)$$

$$\frac{d\sigma_{k0}}{dt} = -2(\eta_{k0} + i2V_{(4)} n_{k0})\sigma_{k0} - i2[V_{(3)} \alpha_0 / p_{k0} + (h\rho)_{\mathit{eff}}]n_{k0} \quad . \qquad (82)$$

Note that in (82) the term representing the coupling with the $k = 0$ mode is divided by the number of modes $p_{k0}$ assumed in the driving because $\sigma_{k0}$ represents only one pair-mode $k_0$. The coupling term in (82) represents a reaction of the $k = 0$ mode that influences the behavior of the BEC modes. In steady-state $d/dt = 0$ (81) leads to,

$$\alpha_0 = \frac{-ip_{k0} V_{(3)}}{\eta_0 + iV_{(4)} n_0} \sigma_{k0} \quad . \qquad (83)$$

This result, valid for the resonance condition $\omega_0 = 2\omega_{k0}$, shows that the BEC magnon pairs drive the uniform mode as an effective microwave magnetic field by means of the three-magnon interaction. Note that there is no threshold condition in this process, BEC magnon pairs with any value of $n_{k0}$ will create $k =$ 0 magnons. This is in contrast to the so-called subsidiary resonance instability process in which the three-magnon splitting process occurs only if the microwave field exceeds a critical value [34,37,45]. The presence of the term $iV_{(4)} n_0$ in the denominator due to the four-magnon interaction represents a detuning from the resonance condition due to the renormalization of the $k = 0$ mode frequency. In fact, this term is responsible for the saturation in the growth of the $k = 0$ mode amplitude with microwave pumping power observed experimentally [6]. In order to compare theory with data we have solved numerically the coupled equations (81) and (82) with their real and imaginary parts to find the steady-state values of the magnon populations $n_0$ and $n_{k0}$ for each value of the pumping power. The calculations were done considering that the relaxation of all modes involved is dominated by the magnetic interactions, $\eta_0 = \eta_{k0} = \eta_m$. We also use normalized variables and parameters: $n'_k = n_k / SN$, $t' = \eta_m t$, $V'_{(3)} = V_{(3)} (SN)^{1/2} / 2\eta_m$, $V'_{(4)} = V_{(4)} (SN) / 2\eta_m$ and $(h\rho)'_{eff} = (h\rho)_{eff} / \eta_m$. With $4\pi M = 1.76$ kG and $\eta_m = 5 \times 10^7$ s$^{-1}$ we have $V'_{(3)} = 219.0$ and $V'_{(4)} = 1240.0$. Figure 8 shows the variation with microwave power of the normalized steady-state values of the populations of the uniform mode $n_0$ and the BEC mode $n_{k0}$ multiplied by the factor $p_{k0}$. Notice that they are both nonzero only for pumping power above the threshold value.

The total power radiated by the uniform magnetization precessing about the static field with frequency $\omega_0$ is given by [44],

$$P = \frac{2 N^2 \Omega^2 \omega_0^4}{3 c^3} (m_x^2 + m_y^2) \qquad (84)$$

where $N$ is the number of spins in the region of emission, $\Omega$ is the volume of the spin unit cell, $c$ is the speed of light and $m_x$ and $m_y$ are the small-signal components of the transverse magnetization. In (84) we have written the volume of the sample as $V = N\Omega$ to stress the dependence of the radiated power on the square of the number of spins. This characterizes superradiance, a term introduced in 1954 by Dicke [46] to designate the type of spontaneous emission of radiation from an assembly of $N$ atoms that has as an intensity proportional to $N^2$ instead of $N$ as in usual situations. This emission requires some kind of quantum coherence in the atomic states, a topic which became understood many years after Dicke's paper was published. The observation of macroscopic superradiance of microwaves in ferromagnetic resonance in YIG was achieved only in the 1970s [43]. The recent experiments of Dzyapko *et al.* [6] constitute the first observation of superradiance originating from a Bose-Einstein condensate.

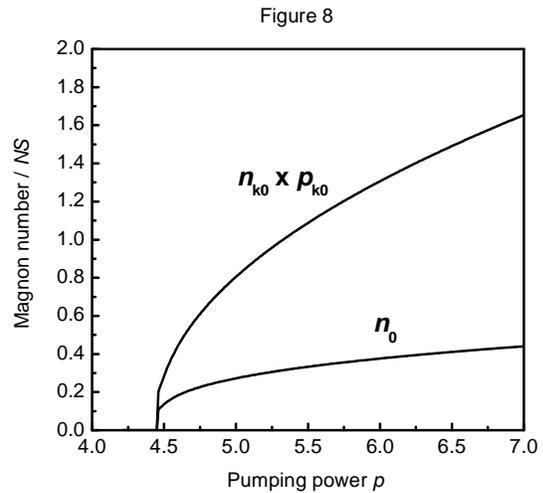

Figure 8: Variation with microwave pumping power of the normalized steady-state magnon numbers of the uniform mode $n_0$ and the BEC mode $n_{k0}$ (multiplied by the factor $p_{k0} = 5 \times 10^4$).

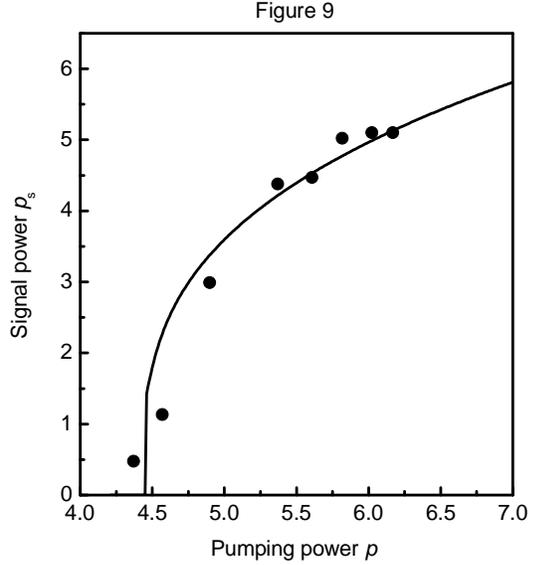

Figure 9: Microwave emission signal power vs pumping power. Symbols represent the experimental data of Dzyapko *et al*. [6] and the solid line is the fit with theory.

Since the microwave signal power is a fraction of the total radiated power given by (84), we use the expression $p_s = C n'_0$ to fit the data of Dzyapko *et al*. [6]. In Figure 9 the symbols represent the data of [6] and the solid line represents the theoretical fit with using $C = 13.2$ μW, $p_{k0} = 5 \times 10^3$ and $p_{c2} = 4.45$ W. The fit is quite good but it is important to check if the values of the fitting parameters bear connection to reality. A good estimate for the number of BEC modes that drive the $k = 0$ magnon is obtained by counting the modes with frequency in the range $\omega_{k0} - \omega_{k0} + \eta_m/2$ and with $\vec{k}$ in the z-direction of the static field, $k_z = n\pi/L_z$, where $n$ is an integer and $L_z$ the sample length. The result obtained numerically with the dispersion relation (26) is $20 \times 10^3$. The value of $p_{k0}$ obtained from the fitting is somewhat smaller than this, which is expected since it represents the number of modes weighted by the number of magnons of the mode $k$ relative to the maximum number at $k_0$. To calculate the emitted microwave signal we use in Equation (84) the expressions for the magnetization components of a coherent state (37) obtaining,

$$\langle P \rangle \cong \frac{V^2 \omega_0^4 M^2}{c^3} n'_0 \qquad . \qquad (85)$$

Using in (85) $\omega_0 = 2\pi \times 3.0$ GHz, $M = 300$ G, $c = 3 \times 10^{10}$ cm/s and an estimated emission volume $V = 1$ mm x 0.5 mm x 5 μm $= 2.5 \times 10^{-6}$ cm$^3$, we obtain for the factor of $n'_0$ in (85) approximately 400 μW. This is two orders of magnitude larger than the value of $C$ obtained from the fit of theory to experiment, which is quite reasonable considering that the measured signal power is only a very small fraction of the total radiated power given by (84). It is important to note that if (81) and (82) are solved considering $p_{k0} = 1$, the calculated $n'_0$ is smaller than the value obtained with $p_{k0} = 5 \times 10^3$ by a factor $10^7$. This means that with $p_{k0} = 1$ the total emitted power calculated with (85) would be smaller than the measured signal power by a factor $10^5$, which is completely unrealistic.

Note that this model is also consistent with the 6 MHz linewidth of the microwave emission spectrum observed in [6]. This value was considered too large by the authors of [6] who expected a linewidth one order of magnitude smaller corresponding to the spin-lattice relaxation rate. In fact the linewidth is close to the value determined by the magnetic relaxation rate, $\eta_m/2\pi = 8$ MHz, which in our theory dominates decay process.

## VIII- Summary

In conclusion, we have shown that in a magnon system in a YIG film driven by microwave radiation far out of equilibrium, the four-magnon interactions acting on the reservoir modes with frequencies close to the minimum in the dispersion relation create the conditions for the spontaneous generation of coherent magnon states. As the microwave power $p$ is increased and exceeds a critical value $p_{c2}$, the magnetic quantum states change from incoherent to coherent magnon states. Correspondingly, the small-signal magnetization changes from zero to $m^+ \propto (p - p_{c2})^{1/4}$ for $p \geq p_{c2}$. Since the magnetization represents the order parameter of the dynamic magnetic system, this characterizes a true second order phase transition with critical exponent 1/4. The spontaneous emergence of quantum coherence [41] caused by a phase-transition and the associated magnetic dynamic order in a macroscopic scale, constitute rigorous theoretical support for the formation of Bose-Einstein condensation of magnons at room temperature, as claimed by Demokritov and co-workers [1-6]. We have also shown that the nearly uniform mode generated by Bose-Einstein condensate (BEC) magnon pairs emits superradiance as a result of the cooperative action of the spins. The theory explains quantitatively recent experimental observations of Dzyapko *et al*. [6] of microwave emission when the driving power exceeds a critical value. The theoretical results fit very well the data for the emitted signal power versus microwave pumping power with realistic parameters.

The author would like to thank Professor Roberto Luzzi of UNICAMP for calling our attention to the recent challenges of BEC of magnons and Professor Sergej Demokritov of University of Muenster for providing important information on the experiments. The author is also very grateful to Professor Cid B. de Araújo for many stimulating discussions and for the Ministry of Science and Technology for supporting this work.